\documentclass{svproc}
\usepackage{url}

\usepackage[latin1]{inputenc}
\usepackage{float}
\usepackage{latexsym}
\usepackage{graphicx}
\usepackage{amssymb}
\usepackage{cool}
\usepackage{dcolumn}
\usepackage{textcomp}
\usepackage{color}
\usepackage{amsmath}
\usepackage{amsfonts}
\usepackage{hyperref}
\usepackage[capitalize]{cleveref}
\usepackage{cite}

\begin{document}
\mainmatter              
\title{Unified Superfluid Dark Sector}
\titlerunning{Unified Superfluid Dark Sector}  
%
\author{Elisa G. M. Ferreira}
\authorrunning{Elisa G. M. Ferreira} 
%
%
\institute{Max-Planck-Institute for Astrophysics, Karl-Schwarzschild-Str. 1, 85741 Garching, Germany, \\
\email{elisagmf@mpa-garching.mpg.de}\footnote{Invited talk at the "XI Symposium Quantum Theory and Symmetry" (QTS-XI), Montreal, QC, Canada, July 2019.}}

\maketitle              

\begin{abstract}
In this talk I present a new model of a unified dark sector, where late-time cosmic acceleration emerges from the dark matter (DM) superfluid framework. We will start by reviewing the dark matter superfluid model and show how it describes the dynamics of DM in large and small scales. Then we will show that if the superfluid consists of a mixture of two distinguishable states with a small energy gap, such as the ground state and an excited state of DM, interacting through a contact interaction a new dynamics of late-time accelerated expansion emerges in this system, without the need of dark energy, coming from a universe containing only this two-state DM superfluid. I will show the expansion history and growth of linear perturbations, and show that the difference in the predicted growth rate in comparison to $\Lambda$CDM is significant at late times. 
\end{abstract}
\section{Introduction}

Our concordance model, the $\Lambda$CDM model, exhibits an outstanding agreement with current large scale cosmological observations \cite{Ade:2015xua,Anderson:2013zyy,Abbott:2017wau}. In this model the present accelerated expansion is described by a cosmological constant, and the dark matter (DM) is described in the hydrodynamical limit as a fluid with negligible pressure and sound speed, with, at most, very weakly interaction with baryonic matter, the Cold Dark Matter (CDM). However, this simple coarse grained description of those components presents some challenges. The cosmological constant is problematic since its smallness is vexing given its radiative instability under quantum corrections~\cite{Padilla:2015aaa}. On small scales, a number of challenges have emerged for this hydrodynamical desxcription of the CDM~\cite{Bullock:2017xww}, with the most striking being the scaling relations like the mass discrepancy acceleration relation (MDAR), which correlates the dynamical gravitational acceleration inferred from rotation curves and the gravitational acceleration due to baryons only ~\cite{McGaugh:2016leg,Lelli:2017vgz}.


There is a debate about the explanation for these curious relations on small scales. Within $\Lambda$CDM model it is claimed that it can be solved by the inclusion of baryonic feedback effects in simulations (see \cite{review} for a review). An alternative is to modify the behaviour of DM on small scales while maintaining the successes of CDM on large scales. Ultra-light fields have emerged as an alternative DM scenario with a different mechanism to explain the dynamics on small scales where DM forms a Bose-Einstein condensate (BEC) or a superfluid in galaxies (for a review \cite{review} of this class of models).  One model that accomplishes that is the DM superfluid \cite{Berezhiani:2015bqa,Berezhiani:2015pia}, where sub-eV mass particles with sufficiently strong self-interactions thermalize and condense in galaxies. On top of that for a certain superfluid equation of state and in the presence of coupling of the DM phonons to baryons, this theory's effective Lagrangian  is similar to the MOND scalar field theory \cite{Bekenstein:1984tv} that leads to a modified dynamics inside galaxies similar to Milgrom's empirical law\footnote{This empirical law states that the total gravitational acceleration $a$ is approximately the Newtonian acceleration $a_{\rm N}$ due to baryonic matter alone, in the regime $a_{\rm N}\gg a_0$, and approaches the geometric mean $\sqrt{a_{\rm N} a_0}$ whenever $a_{\rm N}\ll a_0$.}~\cite{Milgrom:1983ca} known to explain and predict these scaling relations. 

An interesting question is if the late-time cosmic acceleration can also emerge in the DM superfluid framework, as yet another manifestation of the same underlying substance. We show here that it is indeed possible if we consider that the DM is composed by a mixture of two superfluids, which can be  in two different states of the same superfluid, that are in contact and interacting through a contact Josephson-like interaction  \cite{Ferreira:2018wup}, converting one species into the other.  For the phonons that describe the superfluid this interaction appears as an oscillatory potential that drives the late-time acceleration. The unified vision of the dark sector is attractive for its simplicity, given that in this model needs DM in the form of a superfluid alone to describe both the DM behaviour on large and small scales, and the late-time acceleration.

The DM superfluid model and the unified framework present a series of observational consequences \cite{Berezhiani:2015bqa,Berezhiani:2015pia,Berezhiani:2019pzd} that successfully explain some observational challenges in galactic dynamics, cosmological evolution or present new interesting phenomenological consequences.

\section{Review of Dark Matter Superfluid}

Superfluidity is one of the most striking quantum mechanical phenomena on macroscopic scales. It appears in fluids that when brought to very low temperatures, form a Bose-Einstein condensate, now described by a single wave-function of systems coming from the superposition of the de Broglie wavelength of these bosons. The emergent degree of freedom of this collective system has an emergent new dynamics: it flows without friction.

We want to use the  physics of superfluidity to build a model of DM that on central region of galaxies DM condenses forming a Bose-Einstein condensation with a superfluid phase.  The necessary conditions for condensation, assuming weakly coupling, are that de Broglie wavelength $\lambda_{\rm dB} \sim \frac{1}{mv}$ must be larger than the mean inter-particle separation $\ell \sim (m/ \rho)^{1/3}$, and that particle should interact enough to thermalize. The first condition translates to an upper bound on the mass, $m \lesssim (\rho/v^3)^{1/4}$, which for a MW-like galaxy ($M=10^{12} M_{\odot}$) results in $m \lesssim 4.3~{\rm eV}$. The second condition requires that the particles interact strongly.
An axion-like particle that obeys these conditions condenses on the central regions of galaxies, forming a core, which is enveloped by DM particles that are not condensed and behave like CDM having the usual NFW profile. 

After we guaranteed the conditions for DM to condense on galactic scales, we need to describe the evolution of the superfluid. A superfluid is described by a weakly self-interacting field theory of a complex field $\Psi \propto \rho e^{{\rm i} \Theta}$ with global U(1) symmetry. This symmetry is spontaneously broken by the superfluid ground state of a system at chemical potential $\mu$, so that $\Theta=m t+\theta$. At low energy the relevant degrees of freedom are \textit{phonons}, which are excitations of the Goldstone boson $\theta$ for the broken symmetry. The effective theory of phonons must be invariant under the shift symmetry, $\theta \rightarrow \theta + c$, and Galilean symmetry, appropriate for a non-relativistic superfluid. Therefore, its most general form at leading order in derivatives and zero temperature is given by:
\begin{equation}
{\cal L}_{\rm phonons} = P(X)\,;\qquad X= \dot{\theta} -m\Phi - (\vec{\nabla}\theta)^2 / 2m\,,
\label{LT=0}
\end{equation}
where $\Phi$ is the gravitational potential. The equation of state of the superfluid is encoded in the form of $P(X)$, and the phonon sound speed is given by $c_s^2=\frac{P_{,X}}{\rho_{,X}} = \frac{1}{m} \frac{P_{,X}}{P_{,XX}}$. Superfluids are often described by a polytropic equation of state, $P(X) \sim X^n$, corresponding to $P(\rho) \sim \rho^{\frac{n}{n-1}}$. Written in this form, we can describe a standard weakly-coupled superfluid (BEC DM), for $n=2$; for $n=5/2$ this effective theory describes the Unitary Fermi Gas, a gas of ultra-cold fermionic atoms tuned at unitary. 

In the case of the DM superfluid, since we want to reproduce MOND on galactic scales, this corresponds to $n=3/2$, which gives the expected equation of state for MOND, $P \sim n^3$. One extra ingredient is necessary in order to mediate the MOND force, is that the phonons couple to the baryon mas density. The action for the DM superfluid is given by $\mathcal{L}_{\rm DM}=P_{\rm DM}(X)+\mathcal{L}_{int}$ where:
\begin{equation}
P_{\rm DM}(X) =  \frac{2\Lambda (2m)^{3/2}}{3} X\sqrt{|X|} \,, \qquad {\cal L}_{\rm int} = \alpha\Lambda\frac{\theta}{M_{\rm Pl}} \rho_{\rm b}\,.
\label{PDM}
\end{equation}
where $\alpha$ is dimensionless coupling constant. The square-root form also ensures that the Hamiltonian is bounded from below. This phenomenological interaction term breaks shift symmetry softly.

With this description in hand, we can obtain the halo profile. In the center regions of the halo, we have the superfluid region, where phonon gradients dominate\footnote{The phonon effective field theory breaks down for large phonon gradients, like in the vicinity of stars (e.g. in our solar system). For a more detailed see Sec.~5 of~\cite{Berezhiani:2015bqa}.},  the phonon-mediated acceleration matches the deep-MOND expression $a_{\rm phonon} = \sqrt{a_0 a_{\rm b}}$, where $a_{\rm b}$ is the Newtonian gravitational acceleration due to baryons only. The critical acceleration $a_0$ is related to the theory parameters as $a_0 = (\alpha^3\Lambda^2)/M_{\rm Pl}$. The total force experienced by baryons is the sum of the phonon-mediated force, and the Newtonian gravitational acceleration due to baryons and the DM condensate itself. 


\section{Unified Dark Superfluid}

In this section, we are going to generalize the above model to two non-relativistic superfluid species, described in terms of two distinct phonon excitations, each given by a effective Lagrangian like \ref{LT=0}.  For instance, these could represent two distinguishable states of DM with slightly different energies, $\Delta E \ll m$, such as a ground state (represented with subscript $1$) and an excited state ($2$). The theory of the mixture of these two states has a $U\left(1\right)\times U\left(1\right)$ global symmetry, describing particle number conservation of each species separately. We assume that these species have a contact interaction, the simplest and ubiquitous possible interaction, of the form $\mathcal{L}_{\rm int} \propto -\left(\Psi_1^{*}\Psi_2+\Psi_2^{*}\Psi_1\right) / \left|\Psi_1\right|\left|\Psi_2\right|$. At low energies, this translates into a potential for the phonons:
\begin{equation}
V(\theta_2-\theta_1 + \Delta E \,t) = M^4 \left[1+ \cos \left(\theta_2-\theta_1+ \Delta E \,t \right)\right]\,.
\end{equation}
This is the known and well studied in condensed matter systems Josephson or Rabi coupling. Now number density is not conserved alone anymore, $n \simeq P_{1,X_1}+P_{2,X_2}$, but there is the possibility of conversion between species. Consistent with the non-relativistic approximation we assume that $\Delta E \ll m_i$, and that the mass splinting is large in comparison to $\dot{\theta}_2 - \dot{\theta}_1$, so $V(\theta_2-\theta_1 + \Delta E \,t) \sim V(\Delta E \,t)$.

In the non-relativistic approximation, the pressure is given by ${\cal P} = P_1(X_1) + P_2(X_2) - V(\Delta E \,t)$ and the energy density of the superfluids is:
\begin{equation}
 \rho = \underbrace{\frac{1}{2} (m_1 + m_2) n}_{\rho_{+}} +\underbrace{ \frac{1}{2} \Delta E \left(P_{1\,,X_1} - P_{2\,,X_2}\right)}_{\rho_{-}} + V(\Delta E \,t)\,. 
\label{rhogen_Pgen}
\end{equation}
The adiabatic sound speed of each species, governing the growth of perturbations, is $c_{s\, i}^2 = P_{i\,,X_i}/ (m_i P_{i\,,X_iX_i})$, where $P_{i\,,X_i X_i} \geq 0$ to ensure the $c_{s\, i}^2>0$.

With that, the Friedmann equations for a spatially-flat universe can be written, in the non-relativistic approximation as:
\begin{equation}
3H^2 M_{\rm Pl}^2 = \rho_+ + \rho_- + V(\Delta E \,t) \,, \qquad \dot{H}M_{\rm Pl}^2 \simeq -(1/2)\left( \rho_+ + \rho_-\right)\,.
\label{Friedman}
\end{equation}
The energy density $\rho_+$ redshifts like matter and represents the conservation of number density of DM particles, $\rho_-$ evolves under the influence of the potential and the potential term evolves as dark energy. In the case $n=2$,  the BEC DM, $P(X_i)=\Lambda_i^4 \frac{X_i^2}{m_i^2}$, the "$+$" can be thought as the energy density for the sum of the phases and "$-$" for the difference. This can be recast in the canonical variables\footnote{Coming from the diagonalization of the Lagrangian at leading order in $\Delta E/m_i \ll 1$.} representing the two states of the superfluid: $\xi=(1/N)(N_1^2 \theta_1 + N_2^2 \theta_2)$ and $\chi=(N_1N_2/N) ( \theta_1 - \theta_2)$, with $N_i=\Lambda_i^2/m_i$ and $N=\sqrt{N_1^2+N_2^2}$.

\begin{figure}[htb]
\centering
\includegraphics[scale=0.43]{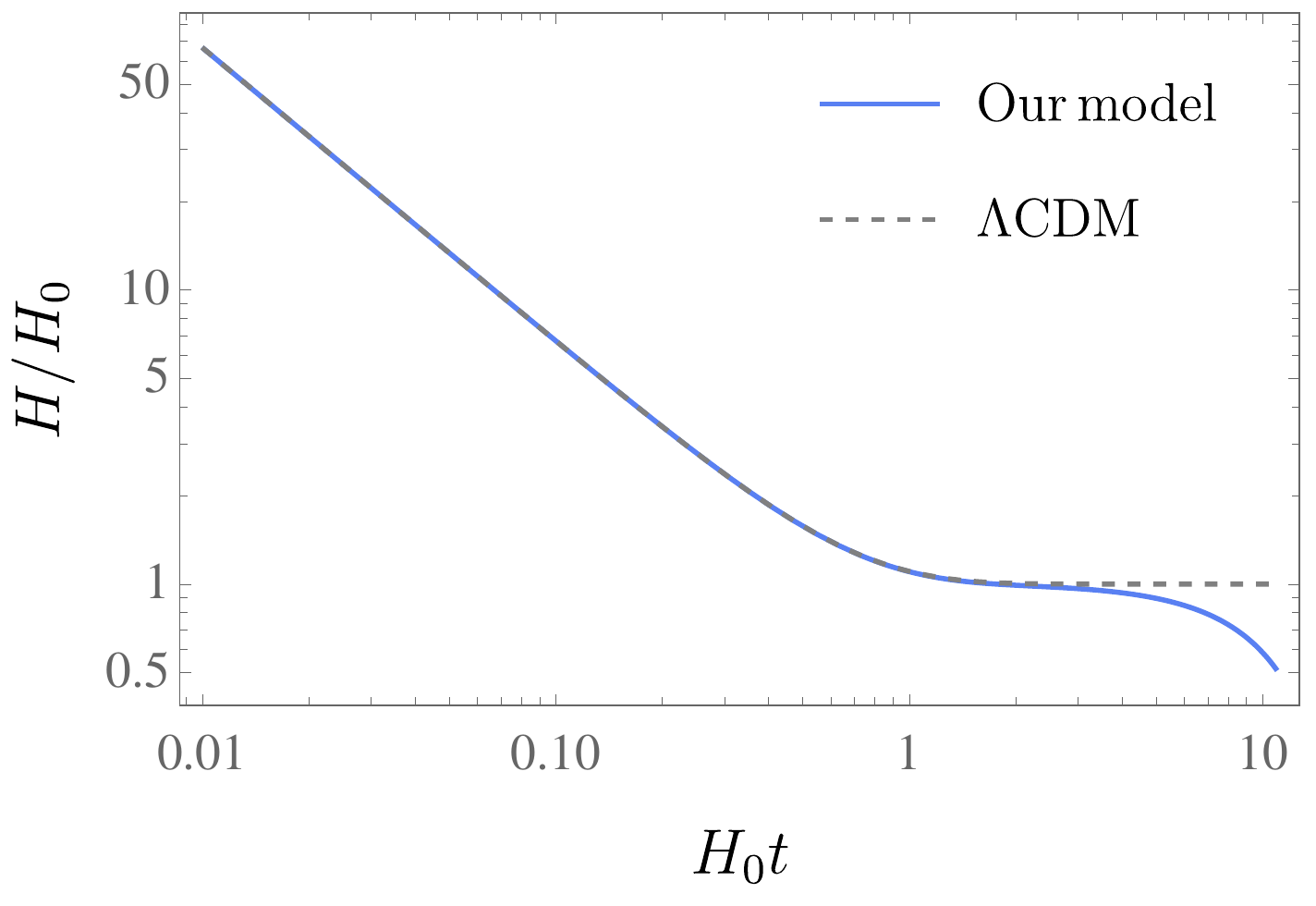}
\caption{Hubble parameter $H$ as a function of time for our model (blue solid curve) for our model in comparison to the $\Lambda$CDM model (black dashed curve).}
\label{fig:Hubble}
\end{figure}

The Friedman equations can be combined leading to a universal equation for the Hubble parameter:
\begin{equation}
2\dot{H} + 3H^2 = V(\Delta E \,t) / M_{\rm Pl}^2\,.
\label{Hclosed}
\end{equation}
From this equation, we can see that during matter domination the potential is not important and the density redshifts as matter; for late times, when the potential becomes dominant, so to ensure that the slow-roll approximation holds for the acceleration we need $\Delta E/ 2H_0 \ll 1$, which implies, for $n=2$, that $\Lambda \gg \sqrt{M_{pl}(m_1+m_2)/2} $ and that the decay constant, the scale of the spontaneous symmetry breaking $f_{\chi}$, is super-Planckian as in the case o pNGB models. 

We can see in Figure \ref{fig:Hubble} the evolution of the unified model in comparison to the concordance model, choosing $M^4 = 2M_{pl}^2 H_0^2 \sim \mathrm{meV}^4$, in order to have acceleration today; and $\Delta E/ 2H_0 = 0.2$.  Our model evolves like $\Lambda$CDM until times close to today, describing the matter era and the late-time acceleration, evolving in a distinct way in the future. 

\section{Growth of Density Inhomogeneities}

A viable alternative to the $\Lambda$CDM model must not only reproduce the evolution of the background, but it should be able to describe the growth of density perturbations that leads to the structures we observe in our universe. In this section we turn to the analysis of density perturbations.

For simplicity we will focus on the BEC DM superfluids. Since our theory describes two interacting superfluids, it is instructive to write down their equations of motion in terms of fluid variables. The continuity and Euler's equations are first-order equations, hence to derive them we must work in the Hamiltonian description. The density is given by $\rho_{\xi}=\Lambda ^2 \Pi_{\xi}$ and $\rho_{\chi}=(\Lambda _1^2 \Lambda _2^2/\Lambda ^2)(\Delta E /m) \Pi_{\chi}$, with $\Pi_{i}=\partial \mathcal{L}/ \partial \dot{i}$ where $i=\xi ,\, \chi$.  Because these were derived in the weak-field approximation, they can be applied to the cosmological context in the free-falling coordinate system (valid for  $H\ell \ll 1$) of the Friedmann-Robertson-Walker (FRW) metric: ${\rm d}s^2 = -(1 + 2\Phi) {\rm d}t^2 + (1-2\Psi) {\rm d}\vec{\ell}^2$, where $\ell$ is the proper distance related  to the coming distance $x$ by the scale factor, $\vec{\ell}=a(t) \vec{x}$. 

Each fluid density can be decomposed into a background piece and an inhomogeneous term: $\rho_\xi = \bar{\rho}_\xi(t) + \delta\rho_\xi (\vec{x},t)$ and $\rho_\chi = \bar{\rho}_\chi(t) + \delta\rho_\chi (\vec{x},t)$.
Note that $\delta\rho_\xi$ and $\delta\rho_\chi$ are {\it not} assumed small at this stage. In this expanding coordinate system, the background densities obey the equations:
\begin{equation}
\begin{array}{l} \dot{\bar{\rho}}_\xi + 3H\bar{\rho}_\xi  = 0
 \\ \dot{\bar{\rho}}_\chi + 3H\bar{\rho}_\chi  = - \Delta E \,V'\left(\Delta E\,t\right)
 \end{array} \,\,\,\, \Longrightarrow \qquad \dot{\bar{\rho}} + 3H\bar{\rho}  = - \Delta E \,V'\left(\Delta E\,t\right)\,,
\end{equation}
where $\bar{\rho} = \bar{\rho}_\xi + \bar{\rho}_\chi$. This confirms, in particular, that $\bar{\rho}_\xi$ describes dust and redshifts as $1/a^3$. Meanwhile, the evolution of $\rho_\chi$ is influenced by the potential.
\begin{figure}[htb]
\centering
\includegraphics[scale=0.34]{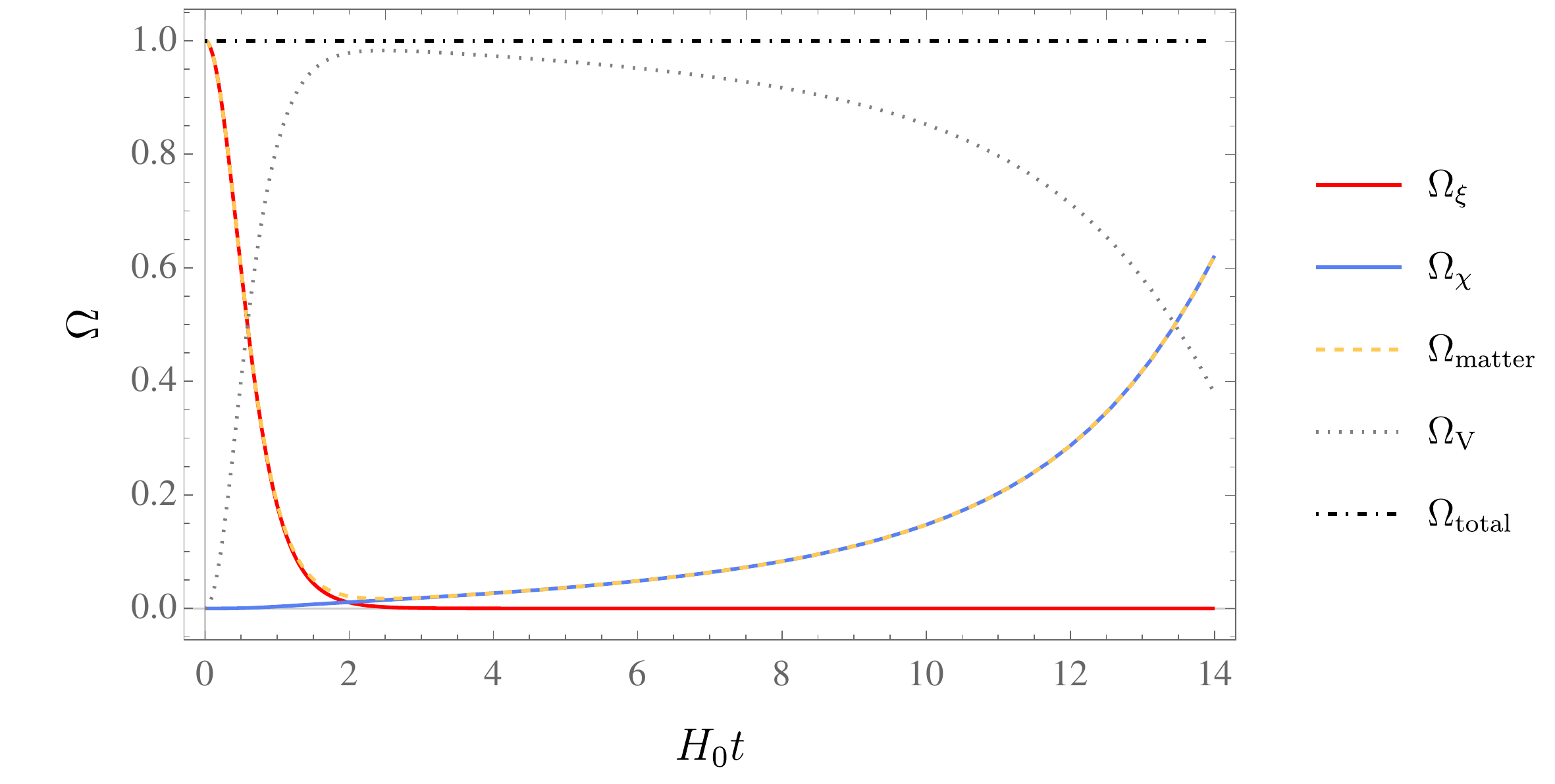}
\caption{Evolution of the fraction density parameters $\Omega_x = \frac{\rho_x}{3M_{\rm Pl}^2 H^2}$ if the components were separated: $\xi$ (red), $\chi$ (blue), the total matter density given by their sum (yellow), and the potential energy (dotted gray).}
\label{fig:densities_future}
\end{figure}
To study the evolution of the background energy density, we solve these equations starting at matter-radiation equality. We will set $m=(m_1+m_2)/2 =  1 \mathrm{eV}$ and $\Lambda_1 = \Lambda_2 = 500 \mathrm{eV}$.   The initial condition for $\bar{\rho}_\xi$ and $\bar{\rho}_\chi$, or for the ground or excited excited state of the superfluid depends on how the energy gap $\Delta E$ compares to the DM temperature at matter-radiation equality, which depends on the production mechanism of our DM particles. For $\Delta E = 5 \times 10^{-11} \mathrm{eV}$, $T_{eq} \sim  10^{-26} \mathrm{eV} \ll \Delta E$, all the matter density is in the ground state $\theta_1$: $\bar{\rho}_\xi^{\;\rm eq} \simeq \rho_{\rm eq} = 0.4~{\rm eV}^4$. In Figure \ref{fig:densities_future} we separate the energy densities of each degree of freedom of the superfluid mixture and the energy density of the potential. The sum of the two superfluid species gives the total DM density.This is only for illustrative purposes, since all of this quantities represent the same fluid. We can see that close to today we have the transition from a matter dominated period to an accelerated expansion, as matter redshifts away and the potential dominates. However, since the potential oscillates and induces conversion of species, eventually the other specie of the superfluid dominates in the future and we will have a new matter domination period. Given the oscillation of the potential, this change might occur many times in the future.

Now, we are interested in analysing the perturbations relative to the total background density, defined as $\delta_i \equiv \delta \rho_i/\bar{\rho}$ with  $i=\xi ,\, \chi$. The fully non-linear equations for the perturbations that describe the Newtonian hydrodynamical equations in an expanding universe can be written as:
\begin{eqnarray}
\nonumber
&& \dot{\delta}_\xi + \frac{1}{a} \vec{\nabla}\cdot \left(\left(\frac{\bar{\rho}_\xi}{\bar{\rho}} + \delta_\xi\right) \vec{v}_\xi\right) - \frac{1}{a} \vec{\nabla}\cdot \left(\left(\frac{\bar{\rho}_\chi}{\bar{\rho}}+\delta_\chi\right) \vec{v}\right) =   \frac{\Delta E \,V'}{\bar{\rho}_\chi}\,\delta_\xi\,;\\
\nonumber
&& \dot{\vec{v}}_\xi  + H \vec{v}_\xi +  \frac{1}{a} \left(\vec{v}_\xi\cdot \vec{\nabla}\right) \vec{v}_\xi  = - \frac{\bar{\rho}}{\Lambda^4} \frac{\vec{\nabla}\delta_\xi}{a} - \frac{\vec{\nabla}\phi}{a}\,;\\
& & \dot{\delta}_\chi + \frac{1}{a} \vec{\nabla}\cdot  \left(\left(\frac{\bar{\rho}_\chi}{\bar{\rho}}+ \delta_\chi\right) \vec{v}_\xi\right)  = \frac{\Delta E \,V'}{\bar{\rho}_\chi}\,\delta_\chi\,; \\
\nonumber
&& \dot{\vec{v}}_\chi  + H \vec{v}_\chi + \frac{1}{a} \left(\vec{v}_\chi\cdot \vec{\nabla}\right) \vec{v}_\chi - \frac{1}{a}\left(\vec{v}\cdot \vec{\nabla}\right) \vec{v} = - \frac{\Lambda^4\bar{\rho}}{\Lambda_1^4\Lambda_2^4}\left(\frac{m}{\Delta E}\right)^2   \frac{\vec{\nabla}\delta_\chi}{a}- \frac{\vec{\nabla}\phi}{a}\,,
\label{fluid_eoms_inhomog}
\end{eqnarray}
where the velocities are given by $\vec{v}_{i}(\vec{l}/a(t),\, t)=\vec{u}_i-H \vec{l}$, with $u_{\xi}=-\vec{\nabla}\xi /\Lambda^2$ and $v_{\chi}=- (\Lambda^2 / \Lambda_1^2 \Lambda_2^2) (m/ \Delta E)\vec{\nabla}\chi$, plus Poisson's equation, 
$\vec{\nabla}^2 \Phi = \frac{a^2}{2M_{\rm Pl}^2} \bar{\rho}\,\delta$.

We can simplify these equations by taking the linear regime, where $\delta_i$ and $v_i$ are small. We can also ignore the spatial gradients,  since both $c_{s, i} \ll 1$. We can then combine the above equations into the equation for the density perturbations:
\begin{equation}
\ddot{\delta} + \left(2H -  \frac{\Delta E \,V'}{\bar{\rho}}\right)\dot{\delta}  =  \frac{1}{2M_{\rm Pl}^2} \bar{\rho}\,\delta + \frac{\Delta E \,V'}{\bar{\rho}}\left(5H +\frac{\Delta E \,V'}{\bar{\rho}}\right) \delta\,,
\label{2nd_order_eoms_final}
\end{equation}
The total velocity evolves as $\dot{\vec{v}}+H\vec{v} \simeq 0$, which redshifts as $1/a$.

\section{Observational Signatures}

Although our model has an evolution that is very close way to $\Lambda$CDM, it predicts distinct observational implications.  We cite some of those in this section.

\begin{figure}[htb]
\centering
\includegraphics[scale=0.35]{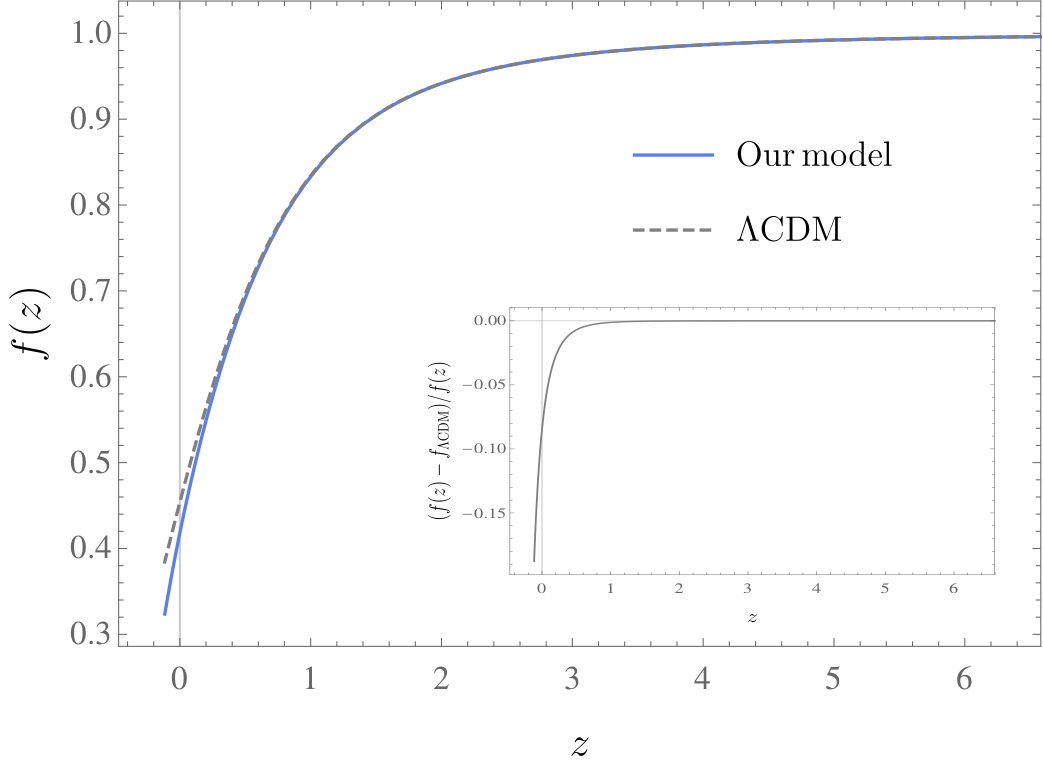}
\caption{Growth rate with respect to the redshift for our model (solid blue line), and the prediction for $\Lambda$CDM (dashed gray line), with initial condition at  equality $\delta_{eq}=10^{-5}$. The fractional difference between our model and $\Lambda$CDM can be seen in the small box. \vspace{-0.5cm}}
\label{growth rate}
\end{figure}

\paragraph{Growth of structures:}
The potential has a distinct evolution than the one of a cosmological constant, affecting also the evolution equations for the density perturbations. This change is explicit in the \textit{growth rate}, $f(z) \equiv - d  \ln \delta (z)/(d \ln (1+z))$, a quantity that is interesting since various probes of structure formation are sensitive, which is shown Figure \ref{growth rate}. We can see that the unified model has a smaller growth rate today than $\Lambda$CDM, which is caused by the potential which is increasing with time, suppressing more structure formation than in $\Lambda$CDM. This difference is around $10\%$ today, for the parameters chosen for the model, which is a difference that can be probed by future experiments.

\paragraph{Vortices:}
Quantum vortices are a prediction of superfluids rotating faster than the critical angular velocity and its measurements would be a smoking gun for the superfluid model. To calculate the abundance and properties of those vortices, it is necessary to have a full microscopic description of the superfluid. 
This topic worth further investigation since the detection of such effect would be an important evidence for the presence of superfluids in galaxies.

For a review of other effects of the DM superfluid in galaxies and clusters, one can see \cite{Berezhiani:2015bqa,review,Berezhiani:2019pzd}
\vspace{0.2cm}

%
%
{\bf \ackname} I would like to thank Justin Khoury, Robert Brandenberger, and Guilherme Franzmann for the most enjoyable and exciting collaboration. I would also like to warmly thank the organizer of the QTS-XI for the stimulating conference and the invitation to speak in the cosmology session.

%
%

\end{document}